\documentclass[seceq]{ptptex}

\usepackage{graphicx}

\notypesetlogo                       

\title{
CFT Description of the Three-Dimensional Hawking-Page Transition
}

\author{
Yasunari \textsc{Kurita}$^{1,}$\footnote{E-mail: kurita@sci.osaka-cu.ac.jp} 
and Masa-aki  \textsc{Sakagami}$^{2,}$\footnote{E-mail: sakagami@phys.h.kyoto-u.ac.jp} %
}

\inst{
$^1$  Department of Physics, Osaka City University, Osaka 558-8585, Japan\\
$^2$  Graduate School of Human and Environmental Studies,
Kyoto University, Kyoto 606-8501, Japan \\
}

\abst{
We construct a phenomenological conformal field theory (CFT) model of 
the three-dimensional Hawking-Page transition. We find that 
free fermion CFT models on the boundary torus give a description of 
the three-dimensional Hawking-Page transition. 
If modular invariance is respected, the free fermion model implies that 
the transition occurs 
continuously through the conical space phase and the small black hole phase. 
On the other hand, if we are allowed to break modular invariance, 
we can construct a free fermion model that reproduces 
the usual Euclidean semi-classical result, 
and in particular exhibits a first-order phase transition. 
}

\begin{document}

\maketitle


\section{Introduction}
\label{intro}

Sometimes, black hole thermodynamics plays one of the most
important roles as a touchstone of quantum theories of gravity.
For the purpose of constructing quantum theories of gravity, 
three-dimensional gravity is important as the first step, 
because there are no local dynamical degrees of freedom, and 
it is relatively easy to discuss its quantum theory.
There is a three-dimensional black hole solution, which is called 
a BTZ black hole \cite{BTZ,BHTZ}. 
It is an asymptotically AdS$_3$ black hole and has almost the same 
thermodynamic properties as higher-dimensional asymptotically AdS black holes. 
This fact implies that local dynamical degrees of freedom 
may not be essential for black hole thermodynamics, and 
it is meaningful to consider three-dimensional gravity  
in order to get an advanced understanding of black hole thermodynamics.

Recently, asymptotically AdS spacetimes have attracted much attention 
because of the AdS/CFT correspondence, which suggests that string theory on
an asymptotically AdS background corresponds to the boundary CFT \cite{Malda,WittenAdS/CFT,GKP}.
Thus it is important to investigate the thermodynamics of asymptotically AdS black holes 
and its CFT counterparts in order to check the AdS/CFT correspondence and to determine how to extract 
information about quantum gravity which might be encoded in the boundary CFT.
In fact, Strominger \cite{Strominger} related the entropy of
a large BTZ black hole to the number of the microscopic states 
in CFT using Cardy's formula \cite{Cardy}.

There is an interesting phenomenon in the thermodynamics of AdS black holes,
the so-called Hawking-Page transition,
which is a transition between thermal AdS space 
and a Schwarzschild-AdS black hole with thermal radiation \cite{Hawking-Page}.
In the three-dimensional case, it is a transition between thermal AdS$_3$ 
and a BTZ black hole.
Some years ago, Witten suggested that 
the Hawking-Page transition corresponds to colour confinement 
in the dual boundary field theory \cite{WittenConfine}. 
Furthermore, the three-dimensional Hawking-Page transition was related 
to the behaviour of correlation functions in the boundary CFT using the AdS/CFT correspondence\cite{BSS2}. 
It was also discussed that 
it can be represented by a transition between gravitational instantons 
in the boundary CFT with modular invariance \cite{MaldacenaStrominger,Mano,BO}. 
The original investigation of the transition by Hawking and Page 
\cite{Hawking-Page} 
was based on the semi-classical approximation of the
Euclidean path integral formulation of black hole thermodynamics. 
However, such an evaluation may not be effective in the vicinity of the transition point.
In the semi-classical evaluation, 
at least two classical solutions, i.e., the AdS$_3$ and BTZ spacetimes,
contribute equally to the partition function at the critical point. 
This fact implies that
fluctuations around the classical solutions have significant effects 
on the transition. Furthermore, some Euclidean solutions interpolating
between them might be important.

In this paper, we attempt to go beyond the semi-classical approximation
by constructing phenomenological boundary CFT models 
on the basis of the asymptotic symmetry
of asymptotically AdS$_3$ spacetime. 
Asymptotic symmetry includes quantum fluctuations of metric fields, and 
the boundary CFT models can be regarded as quantum theories of 
asymptotically AdS$_3$ gravity. 
For this reason, we conjecture that they describe the thermodynamic behaviour, 
even for a range of temperature around the transition point. 
In the next section, we briefly review the three-dimensional 
Hawking-Page transition 
in the Euclidean semi-classical formulation of black hole thermodynamics. 
In \S \ref{CFTdescription}, several CFT models are considered. 
We discuss a free SCFT model in \S \ref{fSCFT}. 
It is found that the SCFT model is not suitable for describing 
the Hawking-Page transition, because its bosonic part gives 
an improper contribution to the internal energy and the specific heat. 
Thus we consider a free fermion model 
as an improved model in \S \ref{ffm}. 
First, we respect the modular invariance in order to construct  
the phenomenological free fermion CFT model.
This model gives a description of the transition in accord with 
the semi-classical result for temperature, apart from the critical one. 
However, in the vicinity of the critical temperature, 
the dominant contribution, which comes only from the NS-NS sector,  
causes the transition to occur continuously between the AdS$_3$ and BTZ
regimes, even in the classical limit.
The model seems to contain contributions from classical solutions with 
a conical singularity which interpolate between AdS$_3$ space 
and a BTZ black hole. Finally, we consider an 
alternative free fermion model 
in \S \ref{no-mod-inv}. 
In this model, half of the fermions are in the NS-NS sector, and the other half 
are in the NS-R and R-NS sectors. It should be emphasized that
this model is not modular invariant, and it exhibits 
much the same behaviour as the Euclidean semi-classical result 
in the classical limit.
At the critical temperature, the internal energy jumps discontinuously, 
so that this model describes a first-order phase transition. 
We give a summary and discuss some issues in \S \ref{summary}.
Finally, in Appendix A we show that the boundary topology of 
asymptotically Euclidean AdS$_3$ spacetime 
is the torus $\it{T}^2$ on which we construct the CFT models in \S 
\ref{CFTdescription}.

\section{Hawking-Page transition in three dimensions}
\label{HP}

In this section, we briefly review the three-dimensional 
Hawking-Page transition, which represents a coexistence of two phases, 
i.e. a BTZ black hole with thermal 
radiation and thermal AdS$_3$.  
We treat this phenomenon in terms of the Euclidean path integral 
formulation of black hole thermodynamics.
Consider a canonical ensemble of a black hole, 
i.e. the situation in which a black hole is in thermal equilibrium with
the radiation at the Hawking temperature $T$.
The partition function of the system is given by a path integral over 
the metric field, which tends asymptotically
to AdS space identified periodically in a Euclidean time $t_E$ with period 
$\beta = 1/T$,
\begin{eqnarray}
Z=\int_{t_{E} \sim t_{E}+\beta} {\mathcal{D}} g
\  e^{-I_{E}[g]}, 
\label{HPpf}
\end{eqnarray}
where $I_E$ is the Euclidean action for gravitational fields with a negative 
cosmological constant.
The partition function for each phase can be evaluated in 
the semiclassical approximation as
\begin{eqnarray}
Z \approx e^{-I_E[\hat{g}]},
\end{eqnarray}
where $\hat{g}$ is the metric for each classical solution, 
the AdS$_3$ and the BTZ black hole. Their classical actions can be calculated 
as follows \cite{BTZ,EJM,Teitelboim}: 
\begin{eqnarray}
I_{E}[\hat{g}_{BTZ}] &=& \frac{1}{8G_3} (\beta M_{BTZ} -4\pi r_+), \\
I_E[\hat{g}_{AdS_3}] &=& -\frac{\beta}{8G_3},
\end{eqnarray}
where $M_{BTZ}$ and $r_+$ are  
the gravitational mass of the BTZ black hole and the radius of 
the event horizon, respectively.
Their contributions to the partition function are 
\begin{eqnarray}
Z_{BTZ}(T) &=& \exp\left(- \frac{\beta M_{BTZ} -4\pi r_{+}}{8G_3} \right)
=\exp\left(\frac{(\pi l)^2 T}{2G_3} \right),
\label{Z_BTZ} \\
Z_{AdS_3}(T) &=&\exp\left(\frac{\beta}{8G_3} \right)=\exp\left(\frac{1}{8G_3T}\right),
\label{Z_AdS3}
\end{eqnarray}
respectively.
In Eq. (\ref{Z_BTZ}), $l$ denotes the AdS raduis.
Then, we obtain the free energies of each spacetime as
\begin{eqnarray}
F_{BTZ}  &=& -\frac{1}{\beta} \log Z_{BTZ} = - \frac{(2\pi l)^2 T^2}{2 G_3}, \label{F_BTZ}\\
F_{AdS_3}&=& -\frac{1}{8G_3}, \label{F_AdS}
\end{eqnarray}  
\begin{figure}[t]
\begin{center}
\scalebox{.6}{\includegraphics{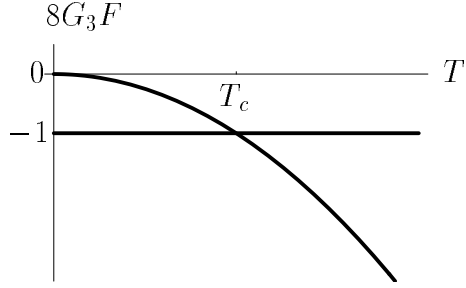}}
\caption{
The free energies of the BTZ black hole $F_{BTZ}$ (\ref{F_BTZ}) and 
the thermal AdS$_3$ background $F_{AdS_3}$ (\ref{F_AdS}). 
Their values coincide at the critical temperature $T_C$.
}
\label{HPF}
\end{center}
\end{figure}
\begin{figure}[htb]
\parbox{\halftext}{
\scalebox{.6}{\includegraphics{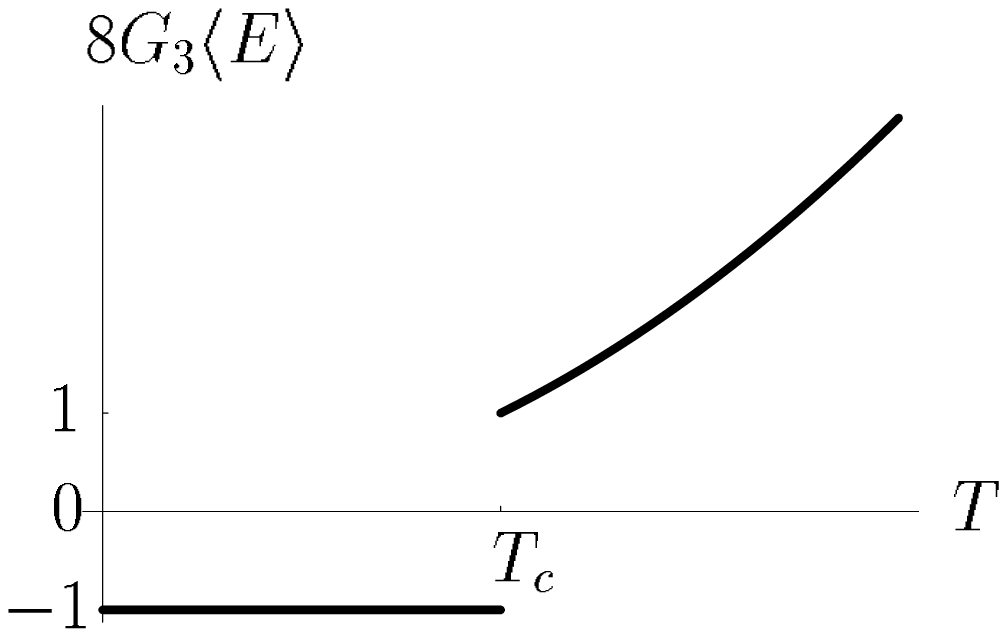}}
\caption{%
The temperature dependence of the energy of asymptotically AdS$_3$ spacetime.
It changes discontinuously at the critical temperature, which means 
that first-order phase transition takes place at $T=T_c$.}
\label{HPE}}
\hfill
\parbox{\halftext}{
\scalebox{.6}{\includegraphics{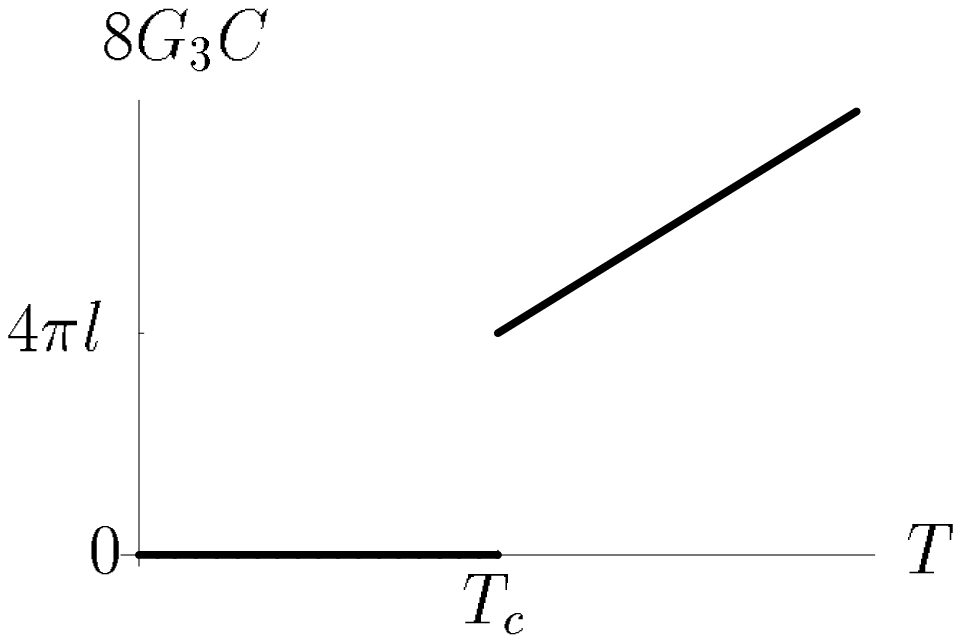}}
\caption{
The specific heat $C$ of the asymptotically AdS$_3$ spacetime 
as a function of $T$. Below the critical temperature, $C$ vanishes. }
\label{HPCv}  }
\end{figure}
with their internal energies 
\begin{eqnarray} 
\langle E\rangle_{BTZ} &=& -\frac{\partial}{\partial\beta}\log Z 
                        = \partial_{\beta}\left(-\frac{4\pi^2 l^2}{8G_3 \beta} \right) =M_{BTZ} 
                        = \frac{(2\pi l T)^2}{8G_3}, 
\label{MBTZ}\\
\langle E \rangle_{AdS_3} &=&-\frac{1}{8G_3} =: M_{AdS_3}, 
\label{MAdS}
\end{eqnarray}
and their entropies  
\begin{eqnarray}
S_{BTZ} &=& \beta \langle E \rangle_{BTZ} + \log Z_{BTZ} 
= 4\pi r_{+}=\frac{A}{4G_3},  \label{semi-entropy} \\
S_{AdS_3} &=& \beta \langle E \rangle_{AdS_3} + \log Z_{AdS_3} = 0.
\end{eqnarray}
We note that the entropy of the BTZ black hole obeys 
the Bekenstein-Hawking law.

As shown in Fig.~\ref{HPF}, the free energy of the thermal AdS$_3$ 
background accords with that of the BTZ black hole  
at the temperatures $T = T_{c} := (2\pi l )^{-1}$. 
For temperatures $T < T_{c}$, the free energy of thermal AdS$_3$ is 
lower than that of the BTZ black hole, so that the BTZ black hole 
is less probable than pure thermal radiation on the AdS$_3$ background. 
On the other hand,  if $T> T_{c}$, the BTZ black hole becomes more probable 
than the pure radiation. At the critical temperature $T=T_c$, 
the two states can coexist. Suppose that we raise the temperature of 
the system gradually from one $T < T_c$. 
Then, obviously, the pure radiation on AdS$_3$ dominates as long as $T$ is 
lower than the critical temperature. When the temperature reaches $T_c$, 
the pure radiation begins to be transformed into the thermal BTZ black hole 
configuration. This is the three-dimensional Hawking-Page transition.
In Fig.~\ref{HPE}, the energy of the asymptotic AdS$_3$ 
spacetime is shown as a function of the temperature. 
We observe a discontinuity at $T=T_c$, which represents the latent heat 
$T_c ~\Delta S = T_c ~(S_{BTZ} - S_{AdS_3})$ 
absorbed during the phase transition. 
Above $T_c$, the state is dominated by the thermal BTZ black hole. 
The specific heat of the asymptotic AdS$_3$ spacetime is also depicted 
in Fig.~\ref{HPCv}. We note that $C$ below the critical temperature
vanishes, since the internal energy for $T < T_c$ is given by 
the constant value $M_{AdS_3}$, as shown in (\ref{MAdS}).

However, the above description of the Hawking-Page transition, which 
relies heavily on the semi-classical approximation of the partition 
function, might not be correct near the critical temperature. 
In other words, the coexistence of two classical solutions contributing to
the partition function at the critical temperature implies the importance 
of fluctuations around these solutions, which have not been taken into account.   
Furthermore, some Euclidean configuration interpolating 
between these two classical solutions or quantum gravitational 
effects should play a significant role. 
Therefore we have to consider other approaches that go beyond the semiclassical one 
in order to investigate the behaviour near the critical temperature.

\section{CFT description}
\label{CFTdescription}

It is well known that asymptotically AdS$_3$ spacetime has a close connection to 
conformal field theory at the boundary surface. 
Because three-dimensional gravity has no local degrees of freedom, 
the boundary theory defined on the two-dimensional surface, i.e. 
the spatial infinity, can describe all of the physics of 
three-dimensional gravity. 
Brown and Henneaux \cite{BrownHenneaux} have shown that 
the asymptotic symmetry of asymptotically AdS$_3$ spacetime 
is Virasoro symmetry with the central charge $c=3l/{2G_3}$, 
which is the two-dimensional conformal symmetry.
This fact strongly suggests that the quantum theory of three-dimensional
AdS gravity can be described by CFTs on the boundary, 
because the asymptotic symmetry also includes quantum fluctuations 
of metric fields.

In this section, we seek CFT models that can give a proper description of the Hawking-Page transition.
We first consider a free superconformal field model as a candidate in \S \ref{fSCFT}. 
Then a modular invariant free fermion model is introduced as an improved model in \S \ref{ffm}.
We also discuss a free fermion model that is not modular invariant in \S \ref{no-mod-inv}.

\subsection{SCFT model}
\label{fSCFT}

Both the AdS$_3$ and massless BTZ spacetimes have global supersymmetries 
at zero temperature associated with AdS$_3$ supergravity, 
and the mass gap between the AdS$_3$ and massless BTZ spacetimes
can be understood as the difference in vacuum energy of 
the boundary CFT with the superconformal algebra \cite{T=0massgap}.
This point was also explained in terms of the AdS/CFT correspondence \cite{BK}.
As a first step, we consider a free supersymmetric CFT (SCFT) model 
in this subsection. Supersymmetry requires that the number of 
bosons be equal to the number of fermions. 
Each free boson contributes $1$ to the central charge, and each fermion contributes $\frac{1}{2}$.
Thus, free SCFT with the central charge $c$ includes $2c/3$  
free bosons and $2c/3$ free fermions.
The boundary topology of asymptotically Euclidean AdS$_3$ is 
a torus $\it{T}^2$ with a modular parameter $\tau = {i}/{2\pi l T}$, 
which is summarized in Appendix \ref{topo}.
The partition function of the boundary CFT is given by 
\begin{eqnarray}
Z_{CFT}(\tau) = 
{\mbox{Tr}}\left[
q^{L_{0}-\frac{c}{24}}\bar{q}^{\bar{L}_{0}-\frac{c}{24}} 
\right],
\label{ZCFT}
\end{eqnarray} 
where $q=\exp (2\pi i \tau)$ and $L_0$ and $\bar{L}_0$ are the left 
and right lowest Virasoro generators.
We can obtain the partition function of the boundary free SCFT as
\begin{eqnarray} 
Z_{SCFT}(\tau)   
&=& (\mbox{Im} \tau)^{-c/3}|\eta(\tau)|^{-2c}\left[
\sum_{i=2}^{4}|\theta_i (\tau)|^{2c/3}
\right],
\label{SCFTparti}
\end{eqnarray} 
which is expressed in terms of the Dedekind eta function $\eta (\tau)$ and 
the elliptic theta functions $\theta_i(\tau)$.\footnote{The definitions and properties of the Dedekind eta and
 elliptic theta functions are listed, for example, in Ref. 19).}
We note that all of these functions are real, because  
the modular parameter $\tau = {i}/{2\pi l T}$ is pure imaginary, 
so that $q=e^{-1/{lT}}$ is real.

In order to observe the relations between the obtained partition function
(\ref{SCFTparti}) and the Hawking-Page transition, let us investigate its
low and high temperature behaviour.
In the low temperature limit, $T\to 0$, in which  
\begin{eqnarray} 
q &\to& 0 ,\qquad \ \ \ \ \ \eta (\tau)\quad \to\  q^{1/24}, \nonumber \\
\theta_2(\tau) &\to& 2q^{1/8},\qquad \theta_3(\tau),\ \theta_4(\tau) \to 1,
\nonumber
\end{eqnarray}
the partition function can be related to that of Ads$_3$ (\ref{Z_AdS3}) as
\begin{eqnarray}
Z_{SCFT}(T) &\to&  2(\mbox{Im} \tau)^{-\frac{c}{3}}q^{-\frac{c}{12}} \nonumber \\
&=&
2(2\pi lT)^{\frac{l}{2G_3}}e^{\frac{1}{8G_3T}} 
=2(2\pi lT)^{\frac{l}{2G_3}}Z_{AdS_3}(T),
\label{PSL}
\end{eqnarray}
where $c=3l/2G_3$ is the central charge of the model. 

Contrastingly,
in the high temperature limit, $T\to \infty\ (q\to 1)$,  
it seems to be difficult to evaluate the partition 
functions (\ref{SCFTparti}). However, the formulae
\begin{eqnarray}
\theta_2(-1/\tau) &=& \sqrt{-i\tau}\theta_4 (\tau), \qquad
\nonumber 
\theta_3(-1/\tau)\ \  =\  \sqrt{-i\tau}\theta_3(\tau), \\
\nonumber 
\theta_4(-1/\tau) &=& \sqrt{-i\tau}\theta_2 (\tau), \qquad
\nonumber 
\eta(-1/\tau)\quad =\  \sqrt{-i\tau}\eta(\tau),
\nonumber 
\end{eqnarray}
which imply the modular $\mathcal{S}$-invariance of the partition function, i.e.,
\begin{eqnarray}
Z_{SCFT}(\tau) &=& Z_{SCFT}(-1/\tau), 
\end{eqnarray}
are helpful for this purpose. 
Let us set $\tilde{q}=e^{-2\pi i/\tau}=e^{-4\pi^2 lT}$. 
In the high temperature limit, it follows that 
$\tilde{q}$ goes to zero and 
the Dedekind eta function and the elliptic theta functions behave as follows:
\begin{eqnarray}
\eta(\tau) &=& (-i\tau)^{-\frac{1}{2}}\eta(-1/\tau) \ \ 
\to (-i\tau)^{-\frac{1}{2}} (\tilde{q})^{\frac{1}{24}}, \nonumber\\
\theta_2 (\tau)&=& (-i\tau)^{-\frac{1}{2}}\theta_4(-1/\tau)\ 
\to (-i\tau)^{-\frac{1}{2}}, \nonumber\\
\theta_3 (\tau)&=& (-i\tau)^{-\frac{1}{2}}\theta_3(-1/\tau)\ 
\to (-i\tau)^{-\frac{1}{2}}, \nonumber\\
\theta_4 (\tau)&=& (-i\tau)^{-\frac{1}{2}}\theta_2(-1/\tau)\ 
\to 2(-i\tau)^{-\frac{1}{2}}(\tilde{q})^{\frac{1}{8}}. \nonumber
\end{eqnarray}
Therefore we can obtain the asymptotic form 
of the partition function as 
\begin{eqnarray}
Z_{SCFT}(T) &\to& 2(\mbox{Im} \tau)^{c/3}(\tilde{q})^{-\frac{c}{12}} 
\nonumber\\
&=& 2 (2\pi lT)^{-\frac{l}{2G_3}}e^{\frac{\pi^2 l^2 T}{2G_3}}
=2(2\pi lT)^{-\frac{l}{2G_3}}Z_{BTZ}(T),
\label{PSH}
\end{eqnarray}
which shows that the most dominant contribution 
comes from $Z_{BTZ}(T)$ at sufficiently high temperatures.

These two limits of the partition function, (\ref{PSL}) and (\ref{PSH}), 
indicate that the Hawking-Page transition might be described 
in terms of the SCFT model, as noted by Mano \cite{Mano}. 
It is clear that the behaviour of the partition function changes 
at the self-dual point under the modular $\mathcal{S}$ transformation, 
i.e. $\tau \to -1/\tau$. 
We can determine the self-dual point as 
\begin{eqnarray} 
\tau =- \frac{1}{\tau}\quad \Leftrightarrow\quad \tau=i \quad 
 \Leftrightarrow \quad 
T=T_c=\frac{1}{2\pi l}. 
\end{eqnarray}
The obtained temperature coincides with the critical temperature 
evaluated in the Euclidean semi-classical approach in \S \ref{HP}. 
Thus, $Z_{SCFT}$ (\ref{SCFTparti}) is expected to be  
the partition function in the canonical ensemble for 
thermal asymptotically AdS$_3$ spacetime.

\begin{figure}[t]
\parbox{\halftext}{
\scalebox{.6}{\includegraphics{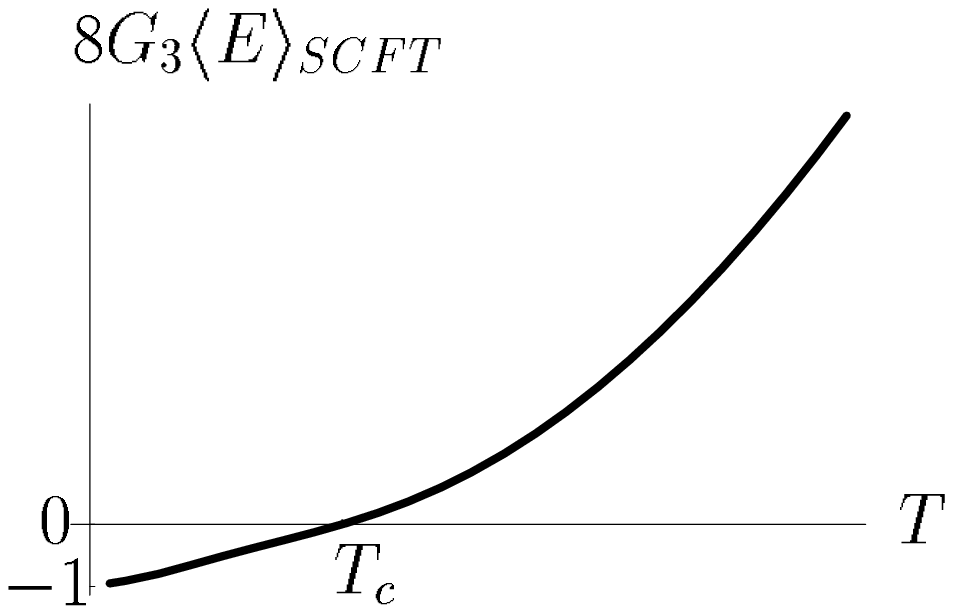}}
\caption{
The temperature dependence of the internal energy  
for the free supersymmetric CFT (SCFT) model given by (\ref{EnSCFT}).
}
\label{E-SCFT}}
\hfill
\parbox{\halftext}{
\scalebox{.6}{\includegraphics{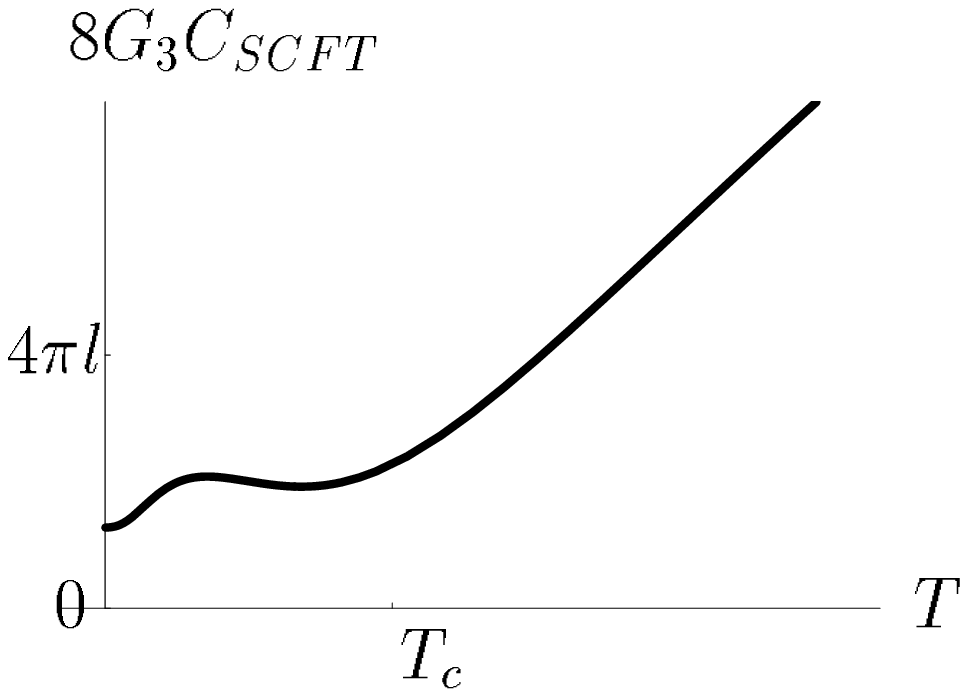}}
\caption{
The specific heat for the SCFT model (\ref{CSCFT}). 
It does not vanish even for zero temperature ($T=0$)
in the classical limit.}
\label{Cv-SCFT} }
\end{figure}

We can calculate thermodynamic quantities from the partition function 
(\ref{SCFTparti}). 
The expectation value of the energy is given by
\begin{eqnarray}
\langle E\rangle_{SCFT} &=& -\partial_{\beta}\log Z_{SCFT}\nonumber \\
 &=& \frac{c}{3}T+\frac{4c}{3}
\frac{\partial_{\beta}\eta(\frac{i\beta}{2\pi l})}{\eta}
-\frac{2c/3}{\sum_{i}|\theta_i|^{2c/3}}
\left[
\sum_{i=2}^4|\theta_i|^{2c/3}
\left(\frac{\partial_{\beta}\theta_i}{\theta_i}-\frac{\partial_{\beta}\eta}{\eta}\right)
\right],
\label{EnSCFT}
\end{eqnarray}
which is plotted in Fig.~\ref{E-SCFT}.
Here, in order to compare the result (\ref{EnSCFT}) with the internal 
energies in the semiclassical approximation given by (\ref{MBTZ}) and (\ref{MAdS}), 
we evaluate it in the low and high temperature cases.
In the low temperature limit, $T \to 0$, 
the derivatives of the eta and theta functions behave as
\begin{eqnarray}
\partial_{\beta}\eta(\tau) &\to& -\frac{1}{24l} q^{\frac{1}{24}},\quad
\partial_{\beta}\theta_2(\tau)\ \ \to\ \  -\frac{1}{4l}q^{\frac{1}{8}}
,\quad \nonumber \\
\partial_{\beta}\theta_3(\tau)&\to& -\frac{1}{l}q^{\frac{1}{2}},\qquad
\partial_{\beta}\theta_4(\tau)\ \ \to\ \  \frac{1}{l}q^{\frac{1}{2}},
\nonumber
\end{eqnarray}
which give the following approximate expression of 
the energy (\ref{EnSCFT}):  
\begin{eqnarray}
\langle E \rangle_{SCFT} &=& \ -\frac{1}{8G_3}+\frac{c}{3}T
+{\mathcal{O}}(e^{-\frac{1}{2lT}}) \nonumber \\
&=&
M_{AdS_3}+\frac{c}{3}T+{\mathcal{O}}(e^{-\frac{1}{2lT}}).
\qquad(T\to 0)
\label{ESCFTl}
\end{eqnarray}
The leading contribution, $-1/8G_3$, is the mass of $\mathrm{AdS_3}$. 
We naively expect the thermal corrections to the leading term 
to be exponentially suppressed in the low temperature regime. 
However, the expression (\ref{ESCFTl}) has a term that is linear with respect to the temperature. 
We note that this term is free from exponential suppression 
and exists even in the classical limit ($G_3 \to 0$, or equivalently, $c\to \infty$).
It is easy to see that only the bosonic part of the partition function
contributes to the unexpected correction.
The high temperature expression of the energy can be obtained  
by means of the modular $\mathcal{S}$-invariance, 
similarly to the case of the partition function.
In the high temperature limit, the derivatives of the eta and 
elliptic theta functions are given by
\begin{eqnarray}
\partial_{\beta}\eta(-1/\tau) &\to&
 \frac{1}{24}(4\pi^2 lT^2)\tilde{q}^{\frac{1}{24}},\quad
\partial_{\beta}\theta_2(-1/\tau)\ \ \to\  \frac{1}{4}(4\pi^2 lT^2)
\tilde{q}^{\frac{1}{8}},\quad \nonumber \\
\partial_{\beta}\theta_3(-1/\tau)
&\to&  (4\pi^2 lT^2) \tilde{q}^{\frac{1}{2}},\qquad
\partial_{\beta}\theta_4(-1/\tau)\quad\to\  - (4\pi^2 lT^2)
\tilde{q}^{\frac{1}{2}}. \nonumber
\end{eqnarray}
Therefore we can obtain the following expression of the energy:
\begin{eqnarray}
\langle E \rangle_{SCFT} &=& 
\ 4\pi^2l^2T^2-\frac{c}{3}T+{\mathcal{O}(T^2 e^{-2\pi^2lT})} \nonumber \\
 &=& M_{BTZ}(T)-\frac{c}{3}T+{\mathcal{O}(T^2 e^{-2\pi^2lT})}.
 \qquad(T\to \infty)
\label{ESCFTh}
\end{eqnarray}
We note that the leading correction to $M_{BTZ}$ is linear 
with respect to the temperature, which comes only from 
the bosonic part of the partition function. 

The specific heat of the system is given by
\begin{eqnarray}
C_{{\tiny SCFT}} &:=& \frac{d}{dT}\langle E \rangle_{SCFT} \nonumber \\
 &=& \frac{c}{3}-\frac{4c}{3T^2}\left(
\frac{\partial_{\beta}^2\eta}{\eta}-\left(\frac{\partial_{\beta}\eta}{\eta}\right)^2
\right) \nonumber
 \\
&\ &
+\frac{2c}{3T^2\tilde{Z}}\left[
\sum_{i=2}^4\theta_i^{2c/3}
\left(
\frac{\partial_{\beta}^2\theta_i}{\theta_i}
-\frac{\partial_{\beta}^2\eta}{\eta}
-\left(\frac{\partial_{\beta}\theta_i}{\theta_i}\right)^2
+\left(\frac{\partial_{\beta}\eta}{\eta}\right)^2
\right)
\right] \nonumber
\\
&\ & +
\frac{2c}{3T^2\tilde{Z}^2}\left[
\tilde{\sigma}_{23}+\tilde{\sigma}_{34}+\tilde{\sigma}_{42}
\right],
\label{CSCFT}
\end{eqnarray}
where we set
\begin{eqnarray}
\tilde{Z}   &:=& \theta_2^{2c/3}+\theta_3^{2c/3}+\theta_4^{2c/3}, \nonumber \\
\tilde{\sigma}_{ij} &:=& (\theta_i\theta_j)^{2c/3}
\left(
\frac{\partial_{\beta}\theta_i}{\theta_i}
-\frac{\partial_{\beta}\theta_j}{\theta_j}
\right). \nonumber
\end{eqnarray}
The temperature dependence of the specific heat is depicted in Fig.~\ref{Cv-SCFT}.
As shown in the previous section, the specific heat evaluated 
using the Euclidean approach vanished at leading order below the critical temperature, i.e., for $T < T_c $. 
Even if we take account of thermal correction, we might have a next-to-leading term 
that is exponentially suppressed by a Boltzmann factor in 
the low temperature region.
In contrast to the result of the Euclidean semi-classical approach, 
the energy $\langle E\rangle _{SCFT}$ has another term which is 
proportional to the temperature, as shown in (\ref{ESCFTl}), so that
the specific heat $C_{SCFT}$ does not vanish even for 
zero temperature  $T = 0$ in the classical limit.
Since the specific heat represents an effective number of 
degrees of freedom at temperature $T$, this deviation of 
$C_{SCFT}$ from that of the semi-classical calculation strongly 
suggests that the SCFT model contains another contribution 
that is unrelated to AdS$_3$. Thus we conclude that 
the SCFT model is not suitable to describe the thermodynamics of the asymptotically AdS$_3$ spacetime.

\subsection{Free fermion model with modular invariance }
\label{ffm}

As shown in the previous subsection, the energy and specific heat evaluated using the SCFT model 
has an extra term which differs from the prediction of the semi-classical approach.
Also, we note that the unfavourable contribution comes only from 
the bosonic part of the SCFT partition function.
Thus, it is quite natural to consider purely fermionic models.  

Since each free fermion contributes $\frac{1}{2}$ to the central charge, 
CFT consisting of $2c$ free fermions on the boundary torus has a central charge $c$. 
Therefore it is also a candidate for the corresponding boundary CFT.
The partition function of this model is given by 
\begin{eqnarray}
Z_{f}(\tau)  &=&
\left(\frac{\theta_2(\tau)}{\eta(\tau)}\right)^{{2c}}
+\left(
\frac{\theta_3(\tau)}{\eta(\tau)}
\right)^{{2c}}+
\left(
\frac{\theta_4(\tau)}{\eta(\tau)}
\right)^{{2c}},
\label{FFparti}
\end{eqnarray}
where $({\theta_2}/{\eta})^{{2c}}$, $({\theta_3}/{\eta})^{{2c}}$ and 
$({\theta_4}/{\eta})^{{2c}}$ are the contributions from R-NS, NS-NS and NS-R sector, respectively.
It should be emphasized that 
the modular invariance of the partition function,
\begin{eqnarray}
Z_f(\tau)\  = Z_f (-1/\tau), \qquad
Z_f(\tau+1) = Z_f(\tau),
\label{ModularTrans}
\end{eqnarray}
is required in the derivation of the partition function (\ref{FFparti}).
Using a method similar to that used in the previous subsection, we can get 
asymptotic forms of the partition function in the low and high temperature limits:
\begin{eqnarray}
Z_f(\tau) &\to & 2q^{-\frac{c}{12}}
\simeq
e^{\frac{1}{8G_3T}}\ =Z_{AdS_3}(T),\qquad (T\to 0) \\
Z_f(\tau) &\to & 2\tilde{q}^{-\frac{c}{12}}
\simeq e^{4\pi^2l^2 T} =Z_{BTZ}(T). \qquad (T\to \infty)
\end{eqnarray}
Furthermore, the critical temperature $T_c$ is also determined by 
the self-dual point 
under the modular $\mathcal{S}$ transformation, as in the SCFT model.

\begin{figure}[htb]
\parbox{\halftext}{
\scalebox{.6}{\includegraphics{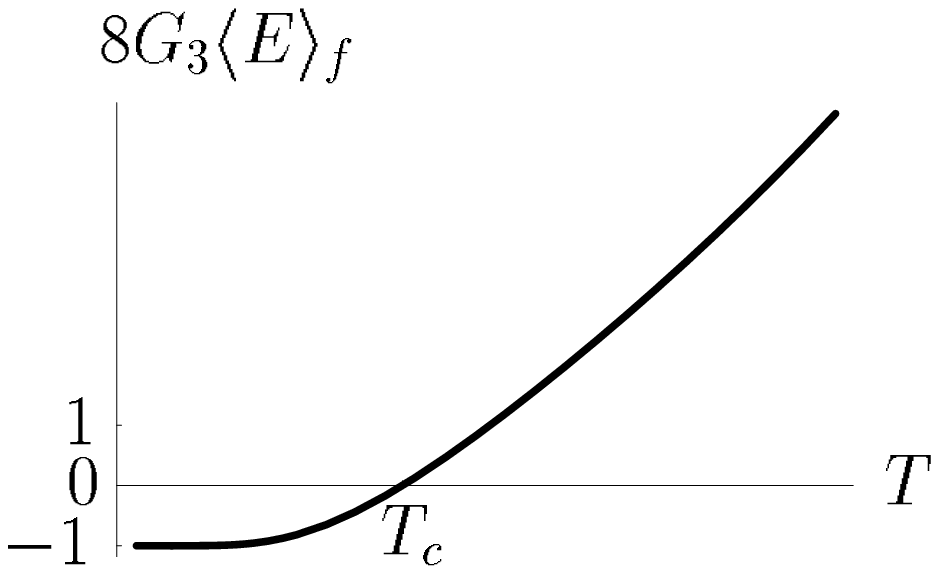}}
\caption{
The temperature dependence of the internal energy for 
the modular invariant free fermion model (\ref{Ef}).}
\label{Efg} }
\hfill
\parbox{\halftext}{
\scalebox{.6}{\includegraphics{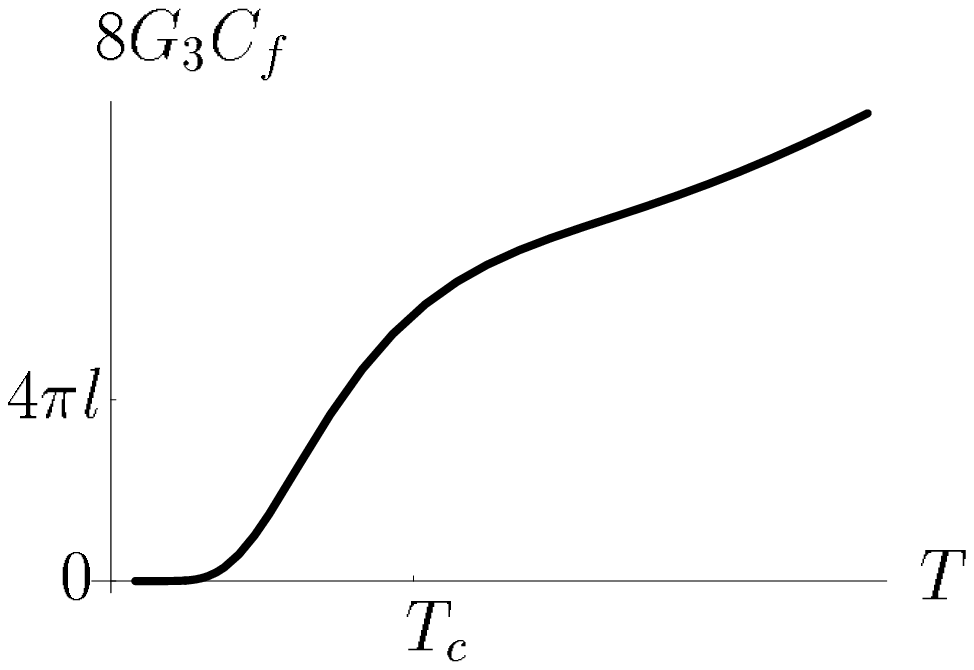}}
\caption{
The specific heat for the modular invariant free 
fermion model (\ref{Cf}). 
In the low temperature limit, $T \ll T_c$, $C_f$ rapidly approaches zero. 
This behaviour is the same as that of the semi-classical result.
}
\label{Cffig} }
\end{figure}
The expectation value of the energy is given by 
\begin{eqnarray}
\langle E\rangle_{f} &=& -\partial_{\beta}\log Z_{f} \nonumber \\
 &=& 
-\frac{2c}{\sum_{i}|\theta_i|^{2c}}
\left[
\sum_{i=2}^4|\theta_i|^{2c}
\left(
\frac{\partial_{\beta}\theta_i}{\theta_i}-\frac{\partial_{\beta}\eta}{\eta}
\right)
\right],
\label{Ef}
\end{eqnarray} 
which is plotted as a function of temperature in Fig.~\ref{Efg}.
In the low and high temperature limits, it becomes
\begin{eqnarray}
\langle E \rangle_{f} &=& \ -\frac{1}{8G_3}
+{\mathcal{O}}(e^{-\frac{1}{2lT}}) \nonumber \\
 &=& M_{AdS_3}+{\mathcal{O}}(e^{-\frac{1}{2lT}}),  \qquad(T\to 0)
\label{EfA1} \\
\langle E \rangle_{f} &=& \  4\pi^2l^2T^2
+{\mathcal{O}(T^2e^{-2\pi^2lT})}  \nonumber\\
 &=& M_{BTZ}(T)+{\mathcal{O}(T^2 e^{-2\pi^2lT})}. \qquad(T\to \infty)
\label{EfA2}
\end{eqnarray}
Unlike the case of the SCFT model, the free fermion model does not contain 
an unfavourable term (i.e., one is proportional to the temperature) 
in these asymptotic expressions (\ref{EfA1}) and (\ref{EfA2}). 
If the temperature is sufficiently lower than the critical one, i.e. $T \ll T_c$, 
the energy 
$\langle E \rangle_{f}$ is approximately equal to the mass of $\mathrm{AdS_3}$ 
space ($M_{AdS_3}=-1/8G_3$), and the thermal correction to it is 
exponentially suppressed by the Boltzmann factor. 
On the other hand, $ \langle E \rangle_{f}$ is dominated by
the mass of a BTZ black hole in the case $T\gg T_c$.
Thus, this model reproduces the Euclidean semi-classical result when
temperature is sufficiently far from the critical point.
The specific heat in this model is given by
\begin{eqnarray}
C_{f} &:=& \frac{d}{dT}\langle E \rangle_{f} \nonumber\\
 &=& \frac{2c}{T^2\bar{Z}(\tau)}\left[
\sum_{i=2}^4\theta_i^{2c}
\left(
\frac{\partial_{\beta}^2\theta_i}{\theta_i}
-\frac{\partial_{\beta}^2\eta}{\eta}-(\frac{\partial_{\beta}\theta_i}{\theta_i})^2
+(\frac{\partial_{\beta}\eta}{\eta})^2
\right)
\right] \nonumber
\\
&\ & +
\frac{2c}{T^2\bar{Z}^2(\tau)}\left[
\bar{\sigma}_{23}(\tau)+\bar{\sigma}_{34}(\tau)+\bar{\sigma}_{42}(\tau)
\right],
\label{Cf}
\end{eqnarray}
where
\begin{eqnarray}
\bar{Z}(\tau)   &:=& \theta_2^{2c}+\theta_3^{2c}+\theta_4^{2c}, \nonumber\\
\bar{\sigma}_{ij}(\tau) &:=& (\theta_i\theta_j)^{2c}
\left(
\frac{\partial_{\beta}\theta_i}{\theta_i}-
\frac{\partial_{\beta}\theta_j}{\theta_j}.
\right). \nonumber
\end{eqnarray}
The behaviour of the specific heat is shown in Fig.~\ref{Cffig}.
It  approximately vanishes in the low temperature case, $T \ll T_c$, 
and grows linearly as a function of the temperature for $T\gg T_c$, 
as in the case of the Euclidean semi-classical approach.
Therefore, the free fermion model is more capable of describing 
the three-dimensional Hawking-Page transition than the free SCFT model.

However, in the vicinity of the critical temperature, $T\approx T_c$,
the behaviour of $C_{f}$ predicted by the free fermion model is not similar to the semi-classical result. 
In the semi-classical treatment of the Hawking-Page transition, 
the energy changes discontinuously at $T = T_c$, as shown in Fig.~\ref{HPE}.
It follows that the asymptotically AdS spacetime exhibits a first-order phase transition at the critical temperature. 
Contrastingly, the energy evaluated using the modular invariant free fermion model does not 
have any discontinuity, as shown in Fig.~\ref{Efg}.  
One might expect that, in the classical limit $G_3 \to 0$, 
the transition in the model becomes sharp and close to the semi-classical result.  
However, this is not the case, as we show below.

\begin{figure}[t]
  \parbox{\halftext}{
 \scalebox{.6}{\includegraphics{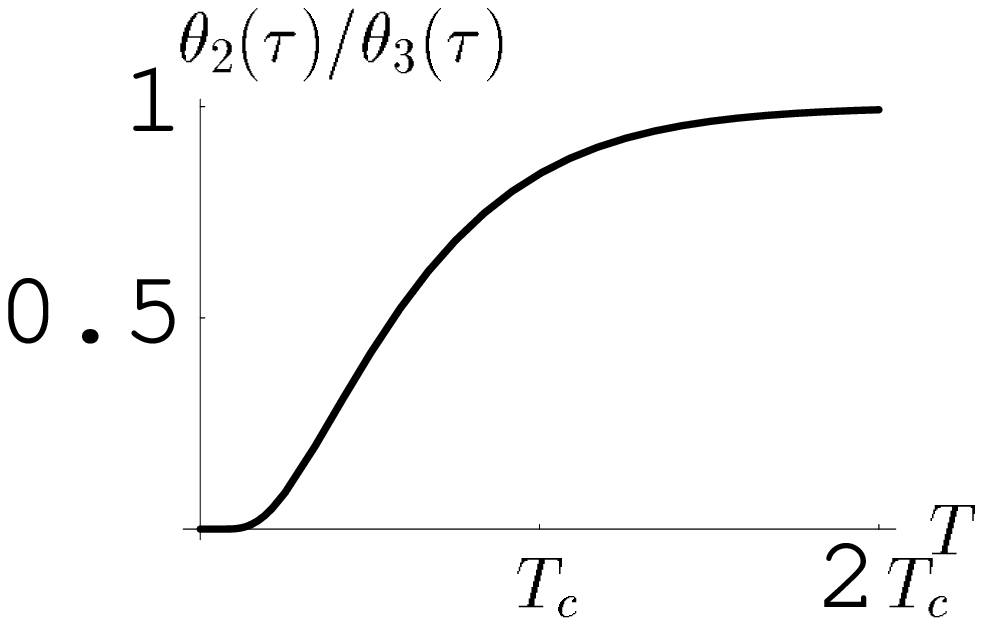}}
   \caption{The temperature dependence of $\theta_2(\tau)/\theta_3(\tau)$. 
This quantity is less than unity at $T=T_c$. } 
\label{2/3} }
\hfill
\parbox{\halftext}{
    \scalebox{.6}{\includegraphics{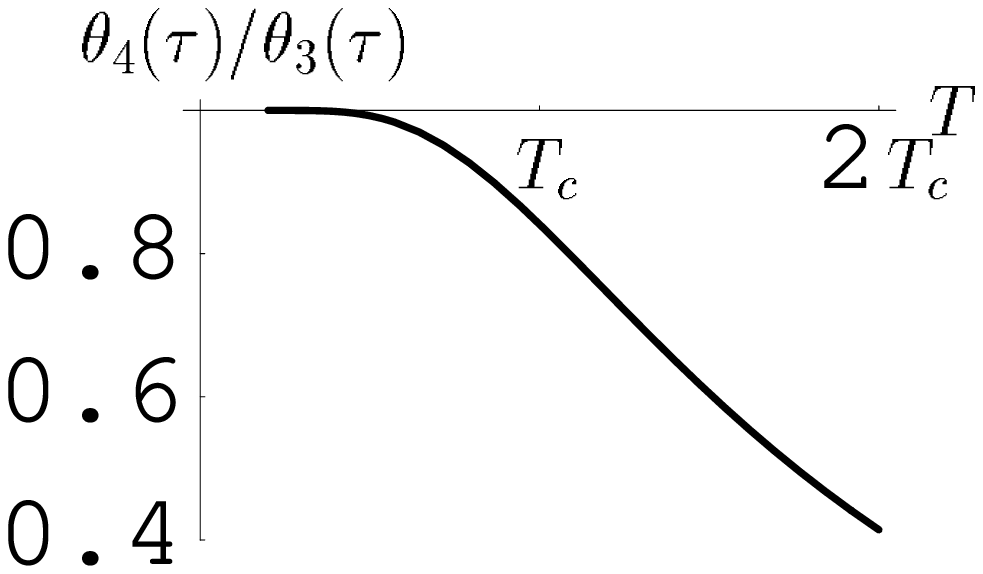}}
\caption{The temperature dependence of $\theta_4(\tau)/\theta_3(\tau)$.
This quantity is less than unity at $T=T_c$. 
}
\label{4/3} }
\end{figure}

\begin{figure}[t]
\begin{center}
\scalebox{.6}
{\includegraphics{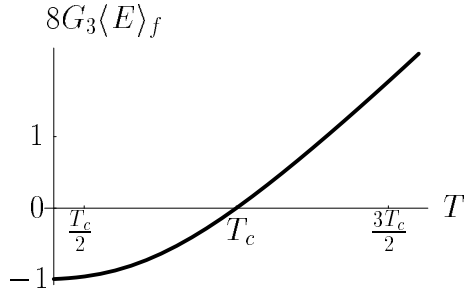}}
\caption{
The temperature dependence of the internal energy (\ref{Efrom3}) 
around the critical point $T_c$ in the case $8G_3=0.01$. 
Even for much smaller values of $G_3$, the internal energy 
varies smoothly and does not exhibit any sign of discontinuity,
which has been observed in the semi-classical approach. 
} 
\label{NSNSenergy}
\end{center}
\end{figure}

In Figs.~\ref{2/3} and \ref{4/3}, through the relation $\tau = {i}/{2\pi l T}$, 
we plot the ratios $\theta_2(\tau)/\theta_3(\tau)$ and $\theta_4(\tau)/\theta_3(\tau)$ 
as functions of the temperature $T$, respectively .  
These figures show that both ratios are less than unity near the critical temperature.
Thus, in the classical limit $G_3 \to 0\ ( c\to \infty)$, these ratios approach zero:
\begin{eqnarray}
\left(
\frac{\theta_2}{\theta_3}
\right)^{2c} \to 0, \qquad
\left(
\frac{\theta_4}{\theta_3}
\right)^{2c} \to 0.
\qquad  (\mbox{for}\quad  T\approx T_c) 
\end{eqnarray}
The energy in the free fermion model (\ref{Ef})
can also be expressed as 
\begin{eqnarray}
\langle E\rangle_{f} &=& 
-\frac{2c}{1+(\frac{\theta_2}{\theta_3})^{2c}
+(\frac{\theta_4}{\theta_3})^{2c}} 
\nonumber \\
&\ & \times
\left[
\left(
\frac{
\theta_3'}{\theta_3}-\frac{\eta'}{\eta}
\right)
+
\left(
\frac{\theta_2}{\theta_3}\right)^{2c}
\left(
\frac{\theta_2'}{\theta_2}-\frac{\eta'}{\eta}
\right)+
\left(
\frac{\theta_4}{\theta_3}\right)^{2c}
\left(
\frac{\theta_4'}{\theta_4}-\frac{\eta'}{\eta}
\right)
\right],
\nonumber
\end{eqnarray} 
where the prime denotes differentiation with respect to 
the inverse temperature $\beta$.
Then, we obtain the asymptotic form of the energy near 
the critical point  in the classical limit as
\begin{equation}
8G_3 \langle E\rangle_{f} \sim -24l
\left(
\frac{\partial_{\beta}\theta_3}{\theta_3}-\frac{\partial_{\beta}\eta}{\eta}
\right), \qquad  G_3 \to 0 .
\label{Efrom3}
\end{equation}
We note that only the NS-NS sector contributes to the energy.  
In other words, the nature of the transition occurring at $T\approx T_c$ is governed by the NS-NS sector rather than
R-NS or NS-R sectors.   
Typical behaviour of the energy (\ref{Efrom3}) in the vicinity of 
the critical temperature is shown in Fig.~\ref{NSNSenergy}.
Even for the case of smaller values of $G_3$, for which 
the system is expected to be close to the classical limit,
the value of the energy changes smoothly from $ -1/8G_3 $ to $ 1/8G_3 $, 
and its behaviour never approaches that of the Euclidean semi-classical treatment. 
We conclude that, although the free fermion model describes a transition between
AdS$_3$ and BTZ spacetimes, this phenomenon cannot be recognized as a phase transition, 
because there is no discontinuity in any thermodynamic quantity.

In the semi-classical approach presented in \S \ref{HP},
the AdS$_3$ and BTZ spacetimes, which are classical solutions 
of the three-dimensional Einstein equation with negative cosmological constant, 
are used for evaluating the partition function of 
the asymptotically AdS spacetime. 
The mass of AdS$_3$ is given by $M_{AdS_3} = -1/8G_3$, and that of BTZ is positive, $M_{BTZ} > 0$, 
so that those two spacetimes exhibit a mass gap. We point out here that 
 for classical solutions satisfying the boundary condition of 
asymptotically AdS, there exists a sequence of spacetimes that have a naked conical singularity 
at the origin and can be interpreted as geometries around a point-like massive source \cite{DJH}. 
Furthermore, it is interesting that the mass of these conical solutions take values $M$ satisfying $-1/8G_3 < M < 0$. 
This range is precisely the mass gap discussed above.
However, this sequence of solutions is not included in the Euclidean semi-classical evaluation of the partition function 
of the asymptotic AdS spacetime in \S \ref{HP}, where only the two classical solutions, 
AdS$_3$ and a BTZ black hole, are taken into account.
These facts imply that the partition function $Z_f$ (\ref{FFparti}) 
calculated using the modular invariant free fermion model contains contributions not only from 
these two classical solutions but also from such  conical spacetimes, 
which  become dominant in the vicinity of the critical temperature and make the transition smooth.

\subsection{Free fermion model without modular invariance}
\label{no-mod-inv}

In the previous subsection, we posed the modular invariance (\ref{ModularTrans}) to the boundary CFT. 
However, in this paper, where we are applying some CFTs for the description of the Hawking-Page transition,
there is no {\it a priori} reason to require the modular invariance of models. 
If modular invariance is not necessary, we are able to consider another model among free fermion CFTs 
which reproduces the result of the Euclidean semi-classical approach.

Let us consider a model consisting of $2c$ free fermions: 
$c$ free fermions exist in the NS-NS sector, and the other $c$ free fermions exist in the NS-R and R-NS sectors. 
It should be emphasized that the NS-NS free fermions are recognized 
as being independent from fermions in NS-R and R-NS sectors.
Then, the partition function is given by 
\begin{eqnarray} 
Z_m(\tau) = \left(
\frac{\theta_3(\tau)}{\eta(\tau)} \right)^c
\left[\left( 
\frac{\theta_2(\tau)}{\eta(\tau)} 
\right)^{c} 
+ 
\left( 
\frac{\theta_4(\tau)}{\eta(\tau)} 
\right)^{c}\right], 
\label{FFMparti}
\end{eqnarray} 
which satisfies the following relation:
\begin{eqnarray}
Z_m(\tau)
= Z_m\left(-\frac{1}{\tau}\right).
\end{eqnarray}
As shown in Appendix \ref{topo}, the modular $\mathcal{S}$ invariance 
$\tau\to -1/\tau$ has the clear meaning that AdS$_3$ spacetime and the BTZ black hole have the same temperature. 
This is the reason why  we construct the partition function (\ref{FFMparti}) to be invariant under 
the modular $\mathcal{S}$ transformation. Contrastingly, this partition function is not invariant 
under the modular $\mathcal{T}$ transformation, $\tau \to \tau +1$, 
which seems to be less important in this case than the modular $\mathcal{S}$ transformation.

The asymptotic forms of the partition function for this model, 
\begin{eqnarray}
Z_m(\tau) &\to& q^{-\frac{c}{12}}
\simeq
e^{\frac{1}{8G_3T}}\ =Z_{AdS_3}(T),\qquad (T\to 0) \\
Z_m(\tau) &\to & \tilde{q}^{-\frac{c}{12}}
\simeq e^{4\pi^2l^2 T} =Z_{BTZ}(T),  \qquad (T\to \infty)
\end{eqnarray}
show that it is dominated by the contribution from AdS$_3$ in the low 
temperature limit and that of BTZ in the high temperature region, as in the previous models. 
The self-dual point of the partition function under the modular 
$\mathcal{S}$ transformation determines the critical temperature $T_c$,
whose value becomes equal to the semi-classical result, as in the previous model.

The energy and the specific heat in this model are given by
\begin{eqnarray} 
8G_3 \langle E\rangle_{m} &= & 
-12l \left( \frac{\theta'_3}{\theta_3}-\frac{\eta'}{\eta} \right) 
 - \frac{12l}{\theta_2^{c}+\theta_4^{c}}
\sum_{i=2,4}
\theta_i^{c}\left(
\frac{\theta_i'}{\theta_i}-\frac{\eta'}{\eta}
\right).
\label{En324m} \\
 8G_3\  C_{m} &=& \frac{12 l}{T^2}\left( \frac{\theta_3''}{\theta_3}
 -\frac{\eta''}{\eta}-\left(\frac{\theta_3'}{\theta_3}\right)^2
 +\left(\frac{\eta'}{\eta}\right)^2 \right)  \nonumber \\ 
 &&+ \frac{12l}{T^2}
 \frac{1}{\theta_2^{c}+\theta_4^{c}}\left[
 \sum_{i=2,4}\theta_i^c
 \left(
 \frac{\theta_i''}{\theta_i}
 -\frac{\eta''}{\eta}-\left(\frac{\theta_i'}{\theta_i}\right)^2
 +\left(\frac{\eta'}{\eta}\right)^2
\right)
 \right] \nonumber \\
 &\ & +12l
 \frac{c}{T^2}\frac{(\theta_2\theta_4)^{c}}{(\theta_2^{c}+\theta_4^{c})^2}
 \left(
 \frac{\theta_2'}{\theta_2}-\frac{\theta_4'}{\theta_4}
 \right)^2.
 \label{Cm324}
\end{eqnarray}
The expressions (\ref{En324m}) and (\ref{Cm324}) clearly show that the internal 
energy and the specific heat are represented as sums of contributions coming from two independent groups of free fermions, 
i.e. those in the NS-NS sector and those in the NS-R and R-NS sectors.
As long as the system is far from the classical limit ($G_3 \to 0$), 
the energy (\ref{En324m}) and specific heat (\ref{Cm324}) of the model behave similarly to those in 
the modular invariant model whose behaviour is shown in Figs.~\ref{Efg} and \ref{Cffig}. 
However, their dependence becomes completely different from 
those of the modular invariant model when the system approaches the classical limit. 

Contrastingly we show that the behaviour of the energy and the specific heat coincides with the results of 
the Euclidean semi-classical analysis discussed in \S \ref{HP}.
\begin{figure}[t]
\parbox{\halftext}{
\scalebox{.6}{\includegraphics{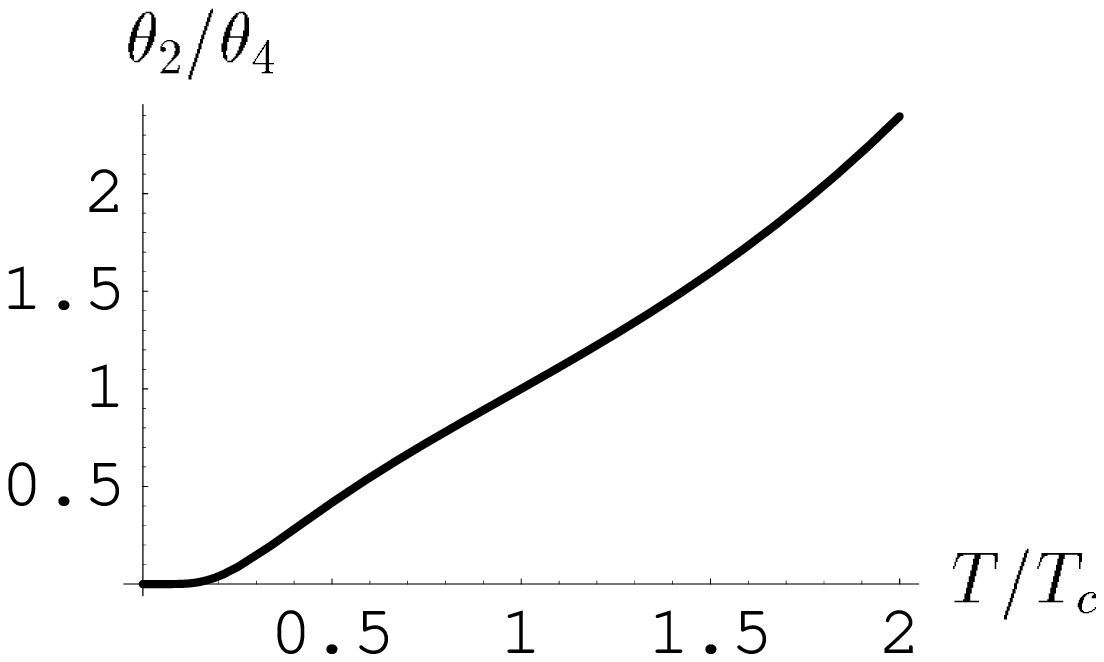}}
\caption{The temperature dependence of the ratio $\theta_2/\theta_4$. 
We see that it is a monotonically increasing function of $T$.}
\label{2-4} }
\hfill
\parbox{\halftext}{
\scalebox{.6}{\includegraphics{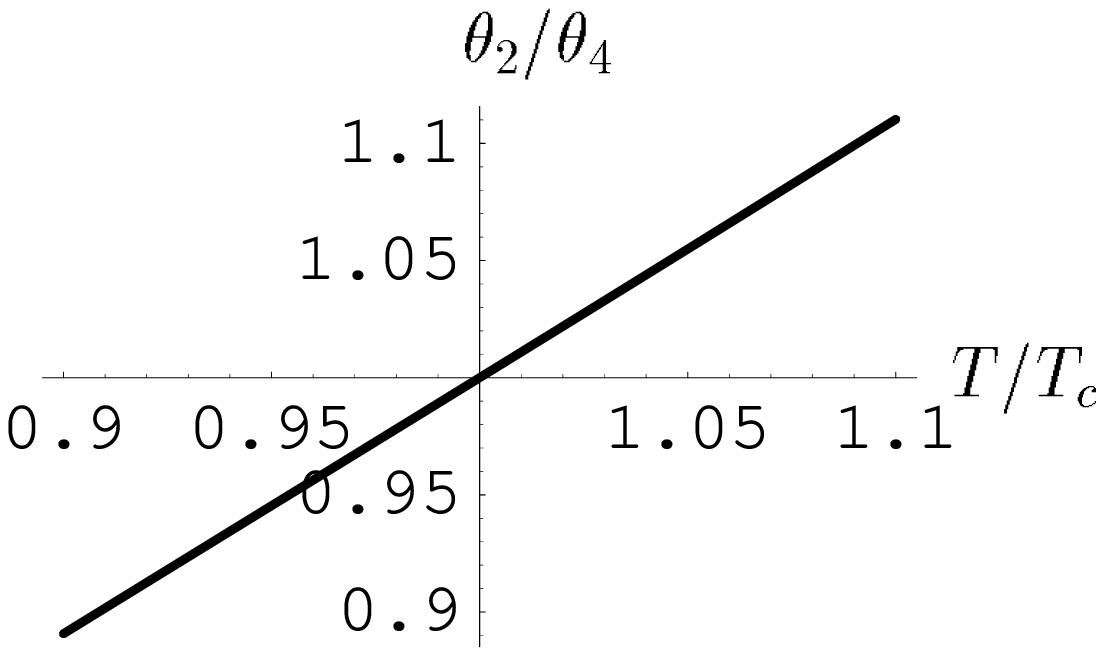}}
\caption{A close-up of the plot of $\theta_2/\theta_4$ 
in the vicinity of the critical temperature. Its value crosses unity at the transition point $T=T_c$. }
\label{2-4-dt} }
\end{figure}
In Figs.~\ref{2-4} and \ref{2-4-dt}, we plot the ratio of $\theta_2(\tau)$ to 
$\theta_4(\tau)$ as a function of the temperature $T$ through the relation $\tau=i/2\pi l T$.
It is  seen that $\theta_2/\theta_4 <1$ for $T<T_c$ and $\theta_2/\theta_4 > 1$ 
for $T>T_c$, so that, in the classical limit ($c\to \infty$), the ratio becomes
\begin{eqnarray}
\left(\frac{\theta_2}{\theta_4} \right)^c  \to
\left\{ \begin{array}{ll}
 0, &  \qquad({\mbox{for}}\ \ T<T_c) \\
 \infty. & \qquad({\mbox{for}}\ \ T>T_c) 
\end{array}
\right.
\end{eqnarray}
Then, we can evaluate the asymptotic form of the energy in 
the classical limit as
\begin{eqnarray}
8G_3 \langle E\rangle_{m} = 
-12l \left( \frac{\theta'_3}{\theta_3}-\frac{\eta'}{\eta} \right) 
 -  12l
\left(
\frac{\theta_4'}{\theta_4}-\frac{\eta'}{\eta}
\right) \quad (T<T_c) 
\label{T<Tc}
\end{eqnarray}
for the low temperature region and 
\begin{eqnarray}
8G_3 \langle E\rangle_{m} = 
-12l \left( \frac{\theta'_3}{\theta_3}-\frac{\eta'}{\eta} \right) 
 - {12l}
\left(
\frac{\theta_2'}{\theta_2}-\frac{\eta'}{\eta}
\right) \quad (T>T_c)
\label{T>Tc}
\end{eqnarray}
for the high temperature region. 
\begin{figure}[t] 
   \begin{center} 
\scalebox{0.6}{\includegraphics{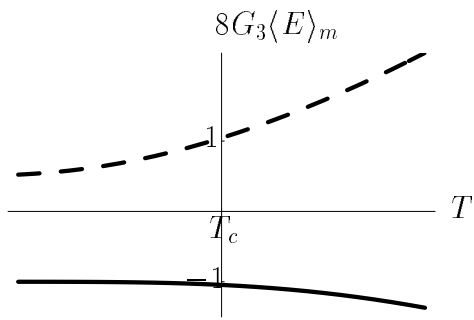}}
\caption{
We plot two expressions for the internal energy of the free fermion CFT without the modular invariance. 
The solid curve represents the formula (\ref{T<Tc}) for the region $T<T_c$, and the dashed curve represents 
(\ref{T>Tc}) for $T>T_c$.
Although both expressions are valid for half of the plotted range of the temperature, 
we depict both for the entire range for illustrative purposes. 
The vertical axis represents the line $T=T_c$. 
Thus, we note that two expressions for the energy have different values at the critical temperature. 
}
    \label{LR}
   \end{center}
\end{figure}
\begin{figure}[ht]  
   \begin{center} 
    \scalebox{0.6}{\includegraphics{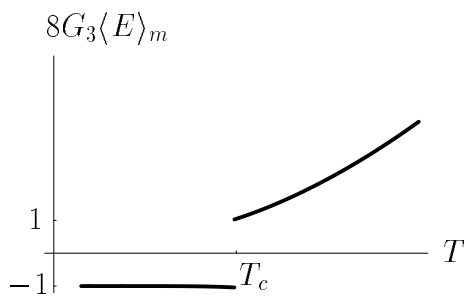}}
\caption{
The solid curve represents the internal energy of the free fermion model without the modular invariance. 
It discontinuously jumps at the transition point.
The expression for the energy is in good agreement with the semi-classical result.
}
    \label{Em}
   \end{center}
\end{figure}
The two functions (\ref{T<Tc}) and (\ref{T>Tc}) are depicted in Fig.~\ref{LR}. 
For illustrative purposes, we have plotted both functions for the whole range of temperatures, 
although these expressions are valid only for restricted regions.
It is clearly noticed that they take different values at the transition point.
This means that the right-side limit ($T\to T_c-0$) of the energy does not agree with the left-side limit ($T\to T_c+0$).  
Hence, the energy changes discontinuously at the transition point. 
Therefore, in the classical limit, the transition described by this model at the critical temperature $T=T_c$ 
is a first-order {\it phase transition}.
In Fig.~\ref{Em}, we show the behaviour of the energy in this model as the solid curve. 
For $T<T_c$, the energy in this model is almost constant and agrees with the mass of AdS$_3$. 
Furthermore, for $T>T_c$, the energy is much the same as the mass of a BTZ black hole.
The behaviour of the energy accords well with the energy in the Euclidean semi-classical approach. 
Thus, if we are allowed to break modular invariance, we can construct a CFT model 
that gives a first-order phase transition and reproduces the Euclidean semi-classical result in the classical limit.

\section{Summary and discussion}
\label{summary}

We have constructed phenomenological CFT models that describe the three-dimensional Hawking-Page transition.
It has been shown that the critical point of the transition corresponds to the self-dual point 
under the modular $\mathcal{S}$ transformation in the boundary CFT. 
Among several candidates, we have found that the SCFT model is not suitable for 
describing the three-dimensional Hawking-Page transition, because it contains another degree of freedom 
which is unrelated to the transition. 
This is consistent with the fact that there is no global supersymmetry in a finite temperature system. 
In contrast to the SCFT model, free fermion CFT models may offer a description of the transition. 
However, unlike the Euclidean semi-classical approach, 
the modular invariant free fermion model exhibits a smooth transition, even in the classical limit. 
In this model, there is no discontinuity in any thermodynamic quantity.  
Therefore, this model does not describe a phase transition. 
Finally, a free fermion model that is not modular invariant has been constructed.  
Among the modular transformations, 
the $\mathcal{T}$ transformation, $\tau \to \tau + 1$, is less significant than the $\mathcal{S}$ transformation, 
$\tau \to - 1/\tau$. For this reason we have posed only $\mathcal{S}$-invariance 
for free fermion models. We have presented one specific model 
which has considerably good agreement with the semi-classical result and exhibits a first-order phase transition. 
Thus, if we are allowed to break modular invariance, we can construct a boundary CFT model 
reproducing the semi-classical result.

In the modular invariant model, the behaviour of the internal energy suggests the existence of some states 
that interpolate between the mass of AdS$_3$ space and that of a BTZ black hole near the transition point. 
In three-dimensional gravity, there is such a sequence of classical solutions with a naked conical singularity, 
which can be interpreted as geometries around a point-like mass source \cite{DJH}. 
Furthermore, it is interesting that the range of the mass of the above conical solutions coincides 
with the mass gap between the AdS$_3$ space and the BTZ black hole. 
This may imply that modular invariance requires the corresponding CFT to include contributions from conical spaces.
Contrastingly, the energy in the modular {\it variant} model jumps discontinuously 
at the critical temperature in the classical limit. 
We speculate that, in the modular {\itshape variant} model, there is no state interpolating between 
thermal AdS$_3$ and a BTZ black hole in its spectrum, and therefore, it behaves like a first-order phase transition. 
In the Euclidean semi-classical argument, we have not included such conical solutions, 
because we do not know how to deal with the conical solutions in the Euclidean formulation of black hole thermodynamics. 
If one could do so, it would be interesting to investigate whether the behaviour of the energy 
in the modular invariant model is reproduced.

In the four-dimensional braneworld scenario, it is known that there are two types of black hole solutions: one is 
the localized black hole solution on an AdS$_3$ brane, and the other is the BTZ black string solution \cite{EHM2}. 
Although the BTZ black string is extended in the bulk space, it looks precisely like a BTZ black hole on the brane. 
Regarding a localized black hole, it should be emphasized that 
its mass ranges from $M=-1/8G_3$ (AdS$_3$) to $M=0$ (massless BTZ).
Thus, it is plausible that the localized black hole solutions correspond to the interpolating 
conical solutions in pure AdS$_3$ gravity, discussed in \S \ref{ffm}, although, in the braneworld scenario, 
their conical singularities can be hidden behind the horizon of four-dimensional black holes. 
Furthermore, the specific heat of a localized black hole is negative, and its free energy is always greater 
than that of the BTZ black string, which has positive specific heat. 
It follows that a localized black hole is thermodynamically unstable \cite{KuKu}. 
These clear correspondences might suggest that we could assign some thermodynamical properties 
even for the conical spaces in pure AdS$_3$ gravity. 
Also, we recognize that the relationship between the conical solution and the BTZ black hole in this paper 
is similar to that in the case of the four-dimensional Hawking-Page transition, 
where there exists a small black hole with negative specific heat, and its free energy is 
always greater than that of a large black hole with positive specific heat.  
We speculate that if we could take into account the contribution from the negative specific heat black hole 
in some sense, the four-dimensional Hawking-Page transition might not be a phase transition,
as suggested by the discussion in \S \ref{ffm}. 

Of course, from the viewpoint of the thermodynamics, there is no room for unstable configurations 
having a significant role in determining how the coexistence of several phases is settled. 
Thus we can conclude that the free fermion model without modular invariance presented in \S \ref{no-mod-inv} should be 
regarded as the most reliable model to describe the three-dimensional Hawking-Page transition. 
As shown in the present paper, it can accurately reproduce the results of the semi-classical Euclidean 
evaluation in the classical limit. 
Furthermore, we recognize that this model should be very useful, since it is a very simple tool for investigating 
how quantum fluctuations affect the nature of the Hawking-Page phase transition in the vicinity of the critical point.  
In the above argument, the modular invariant free fermion model considered in \S \ref{ffm} seems to be less important. 
However, we expect that this is not the case if we are interested in the dynamical aspects of the phase transition, 
e.g., the formation and evolution of domain walls. 
In these situations, similarly to typical approaches employing the Ginzburg-Landau equation, 
we should include effects of fluctuations that permit transitions through unstable configurations 
which have larger free energies than stable configurations.   
In the case of the three-dimensional Hawking-Page transition,
we hope to be able to investigate its dynamics and evolution in terms of the modular invariant free fermion CFT, 
which seems to contain the effects of unstable configurations.

We have not yet discussed the entropy of a BTZ black hole. In the free fermion model, 
the energy of the system is almost the same as that of a BTZ black hole for temperatures $T\gg T_c$.
Then the entropy of the system is given by the thermodynamic relation (\ref{semi-entropy}), 
which is the same as the Bekenstein-Hawking law for sufficiently high temperature.
This is not a microscopic derivation, as in stringy D-brane black holes \cite{StromingerVafa}
or in Strominger's argument \cite{Strominger}. 
Strominger's argument is based on the Cardy formula, which depends on modular invariance \cite{Carlip}. 
Thus we cannot apply that argument to a boundary CFT that is not modular invariant. 
In the Chern-Simons theory, it was shown how unitarity and the demand of the modular invariance of the boundary CFT 
seem to overconstrain the entropy counting problem \cite{TroostTsuchiya}. 
Thus, it is meaningful to consider boundary CFT without modular invariance in this context. 
As an example of such a boundary CFT, let us consider the model discussed in \S \ref{no-mod-inv}. 
Its energy behaves like that of a BTZ black hole for sufficiently high temperature, 
and the thermodynamic entropy of the model obeys the Bekenstein-Hawking law through the relation (\ref{semi-entropy}).
Thus, if we can count the number of physical states in the modular {\it{variant}} model, 
we believe that the result should reproduce the Bekenstein-Hawking formula.

We have restricted our attention to the three-dimensional case in this paper.
The original Hawking-Page transition is concerned with four-dimensional asymptotically AdS spacetime, 
and Witten showed that it exists in higher-dimensional AdS gravity \cite{WittenConfine}.
In higher-dimensional cases, the argument is also based on the semi-classical approximation of the 
Euclidean formulation of black hole thermodynamics, and such an evaluation may not be valid around the critical temperature, 
as in the three-dimensional case. From the standpoint of the AdS/CFT correspondence, 
we conjecture that the corresponding CFT describes the transition even around the critical point. 
Furthermore, we conjecture that the corresponding CFT describes the transition in the manner discussed in this paper: 
The partition function of the corresponding CFT on the Euclidean boundary can be
identified with the partition function of the Euclidean AdS gravity, and 
the CFT partition function gives thermodynamic quantities of the thermal AdS gravity, as in the three-dimensional case. 
Thus, we believe the three-dimensional models considered in this paper may be helpful to understand 
how the boundary CFT describes the Hawking-Page transition in higher-dimensions. 
It is important to consider a generalization of this CFT description to higher-dimensional cases, 
and we leave these issues for future investigations.


\section*{Acknowledgments}
Y.K. would like to thank Tsuneo Uematsu,  Takashi Okamura and 
Shunsuke Teraguchi for useful discussions.
This work was supported by a Grant-in-Aid from the Japan Society for the
Promotion of Science (No. 15540368).


\appendix

\section{Topology of the Asymptotic Boundary}
\label{topo}

In this appendix, we investigate the topology of the boundary
of asymptotically Euclidean AdS$_3$ spacetimes. 
The same argument is given in Refs. \cite{MaldacenaStrominger}and \cite{Mano}.

The metric of a Euclidean non-rotating BTZ black hole is 
\begin{eqnarray}
ds_E^2=(-M+r^2/l^2)dt_E^2+\frac{dr^2}{(-M+r^2/l^2)}+r^2d\phi^2.
\end{eqnarray}
In order to investigate the boundary, we introduce the following coordinates:
\begin{eqnarray}
y &:=& \frac{r_+}{r}\exp \left[\pi l T \phi \right], \\
z&:=&\left(\frac{r^2-r_+^2}{r^2}\right)\exp\left[
2\pi T(l\phi+it_E)
\right].
\end{eqnarray}
Then the Euclidean metric can be expressed in these coordinates as
\begin{equation}
ds_E^2=\frac{l^2}{y^2}\left(
dzd\bar{z}+dy^2
\right).
\end{equation}
In the asymptotic region $r\to \infty$, these coordinates behave as $y\to 0$ and 
\begin{equation}
z\to e^{2\pi T(l\phi+it_E)}.
\end{equation}
In these coordinates, the periodicity of the Euclidean time, $t_E \sim t_E+\beta$, is satisfied automatically, 
but the angular periodicity $\phi\sim \phi+2\pi$ causes $z$ to be periodic:
\begin{equation}
z\sim z\times e^{4\pi^2 lT}.
\end{equation}
We introduce the boundary coordinate 
\begin{eqnarray}
w:=\frac{i}{2\pi}\log z =-T t_E+il T \phi,
\end{eqnarray}
which has two periodicities,
\begin{eqnarray}
w \sim w+1, \qquad
w \sim w+\tau_{\tiny{BTZ}}, 
\end{eqnarray}
where
\begin{eqnarray}
\tau_{\tiny{BTZ}}:=2\pi l T i.
\label{tauBTZ}
\end{eqnarray}
This means that the topology of the asymptotic boundary of a Euclidean BTZ black hole is
a torus $\it{T}^2$ with modular parameter $\tau_{\tiny{BTZ}}$.

Now, note that the metric of Euclidean AdS$_3$ with temperature $T$ is 
\begin{equation}
ds_E^2=\left(1+\frac{r^2}{l^2}\right)dt_E^2
+\left(
1+\frac{r^2}{l^2}
\right)^{-1}dr^2+r^2d\phi^2,
\end{equation}
where the Euclidean time has periodicity $t_E\sim t_E + \beta$.
We introduce the following asymptotic coordinates of this Euclidean space: 
\begin{eqnarray}
\tilde{y} &:=& \left(
\frac{l^2}{r^2+l^2}
\right)^{1/2}e^{\frac{t_E}{l}}, \qquad
\tilde{z} := \left(
\frac{r^2}{r^2+l^2}
\right)^{1/2}e^{\frac{t_E}{l}-i\phi}.
\end{eqnarray}
Then the metric becomes 
\begin{equation}
ds_E^2=\frac{l^2}{\tilde{y}^2}\left(
d\tilde{z} d\bar{\tilde{z}}+d\tilde{y}^2
\right).
\end{equation}
In the asymptotic region $r\to \infty$, these coordinates behave as $\tilde{y}\to 0$ and
\begin{equation}
\tilde{z} \to e^{\frac{t_E}{l}-i\phi}.
\end{equation}
In this case, the angular periodicity $\phi\sim \phi+2\pi$ is trivial, 
but the periodicity of the Euclidean time causes the coordinate $\tilde{z}$ to be periodic in the other direction:
\begin{equation}
\tilde{z} \sim \tilde{z} \times e^{1/T l}.
\end{equation}
We introduce the boundary coordinate for Euclidean AdS$_3$: 
\begin{equation}
\tilde{w} :=\frac{i}{2\pi}\log \tilde{z} 
=\frac{\phi}{2\pi}+i\frac{t_E}{2\pi l}.
\end{equation}
This complex coordinate has the following periodicity:
\begin{eqnarray}
\tilde{w} \sim \tilde{w}+1, \qquad
\tilde{w} \sim \tilde{w}+\tau_{\tiny{AdS}},
\end{eqnarray}
where
\begin{eqnarray}
\tau_{\tiny{AdS}}:=\frac{i}{2\pi l T}. 
\label{tauAdS3}
\end{eqnarray}
Hence, the topology of the boundary of thermal AdS$_3$ is
also a torus $\it {T}^2$ with a modular parameter $\tau_{\tiny{AdS}}$.
The modular parameter of the torus is related to that of the boundary BTZ torus as 
\begin{equation}
\tau_{\tiny{AdS}}=-\frac{1}{\tau_{\tiny{BTZ}}}.
\label{mod-t}
\end{equation}
This relation means that these two modular parameters are related to each other 
by the modular $\mathcal{S}$ transformation ($\tau \to -\frac{1}{\tau}$), 
which corresponds to an interchange of the coordinates $\phi$ and $t_E$.
Therefore, the boundary of the Euclidean BTZ black hole and that of thermal $\mathrm{AdS_3}$ 
with the same Hawking temperature $T$ are always the same torus. 
Thus we can use either modular parameter as that of the boundary torus for asymptotically Euclidean $\mathrm{AdS_3}$, 
as long as we respect the modular $\mathcal{S}$ invariance.


\end{document}